\documentclass[onecolumn]{aastex631} 

\usepackage{xspace} 
\usepackage{float} 

\usepackage[skip=10pt plus1pt, indent=20pt]{parskip} 
\newcommand{\Msun}{\ \rm M_{\odot}}

\begin{document}

\title{\huge A Strong Gravitational Lensing Model of         
PSZ2\,G118.46+39.32}
\author[0009-0008-8293-9404]{Isaac Smith}
\affiliation{Department of Astronomy, University of Michigan 
1085 South University Avenue 
Ann Arbor, MI 48109, USA}
\author[0000-0002-8261-9098]{Catherine Cerny}
\affiliation{Department of Astronomy, University of Michigan 
1085 South University Avenue 
Ann Arbor, MI 48109, USA}
\author[0000-0002-7559-0864]{Keren Sharon}
\affiliation{Department of Astronomy, University of Michigan 
1085 South University Avenue 
Ann Arbor, MI 48109, USA}
\author[0000-0003-3266-2001]{Guillaume Mahler}
\affiliation{STAR Institute, Quartier Agora - All\'ee du six Ao\^ut, 19c B-4000 Li\`ege, Belgium}
\author[0000-0002-3475-7648]{Gourav Khullar}
\affiliation{Department of Astronomy, University of Washington, Physics-Astronomy Building, Box 351580, Seattle, WA 98195-1700, USA}
\author[0000-0002-0443-6018]{Benjamin Beauchesne}
\affiliation{Centre for Extragalactic Astronomy, Department of Physics, Durham University, South Road, Durham DH1 3LE, UK}
\affiliation{Institute for Computational Cosmology, Department of Physics, Durham University, South Road, Durham DH1 3LE, UK}
\author{The SLICE Collaboration}

\date{\today}

\begin{abstract}

\noindent We present the first strong gravitational lensing model for the cluster PSZ2\,G118.46+39.32 ($z = 0.3967$) using new NIRCam imaging from the Strong LensIng and Cluster Evolution (SLICE) JWST program. We leverage the broad coverage of the SLICE ultrawide JWST filters to identify new lensed galaxies, some of which are not visible in HST, to model the cluster’s mass distribution. The model was constructed with a total of 11 multiply imaged systems, decomposed into 30 images with 60 clumps used as strong lensing constraints. PSZ2\,G118.46+39.32 shows a clear bimodal structure, indicating that it may be undergoing a merger. The predicted mass distribution of the model aligns with the X-ray gas in the cluster, suggesting it is in a pre-merger state. 

\end{abstract}

\section{Background} \label{sec: background}

    Cluster mergers present strong evidence for the presence of dark matter (e.g., ``The Bullet Cluster'', \citealt{Clowe2006}). Strong lensing models paired with X-ray observations provide a unique probe into the behavior of dark matter in merging clusters \citep{2024Beauchesne}. The bimodal structure of PSZ2\,G118.46+39.32 (J1354+7714) makes it a good candidate for such analysis. 

\section{Strong Lensing Analysis} \label{sec: analysis}

    We used NIRCam/F150W2 and F322W2 imaging from the Strong LensIng and Cluster Evolution (SLICE; \citealt{Cerny2025}) JWST Cycle 3 Survey program (PID:5594, PI: Mahler), and archival HST ACS/F814W and F606W imaging (Figure~\ref{fig: lensing model}). The cluster was modeled using the publicly available parametric lens modeling software \texttt{Lenstool} \citep{Jullo2007}, which uses Markov Chain Monte Carlo sampling to explore the parameter space and find a foreground mass distribution that minimizes the separation between the observed and predicted image positions.
    
    The lens was modeled as a linear combination of Pseudo-Isothermal Ellipsoidal Mass Distribution (PIEMD, or dPIE; \citealt{2007Eiasdottir}) halos. 195 cluster member galaxies were selected by their color relative to the red sequence \citep{GladderYee2000} in the F606W-F814W color-magnitude diagram. Their morphological parameters ($x$,$y$,$e$,$\theta$) were fixed to their observed values as measured by \textsc{SExtractor} \citep{1996A&AS..117..393B}. The remaining slope parameters were fitted using the scaling relation described in \citealt{1976FaberJackson}. 

    We modeled each sub-cluster with one cluster-scale dark matter halo. We centered one free galaxy-scale halo on the brightest cluster member galaxy (BCG) at the core of the east sub-cluster to account for the complex strong lensing in this region. Another galaxy-scale halo (centered on ``OG" in Figure~\ref{fig: lensing model}) was optimized to account for its significant lensing effect on system 3. Our model has 45 free parameters. 

\section{Strong Lensing Constraints} \label{sec: constraints}

\begin{figure}
    \centering
    \includegraphics[width=0.9\textwidth]{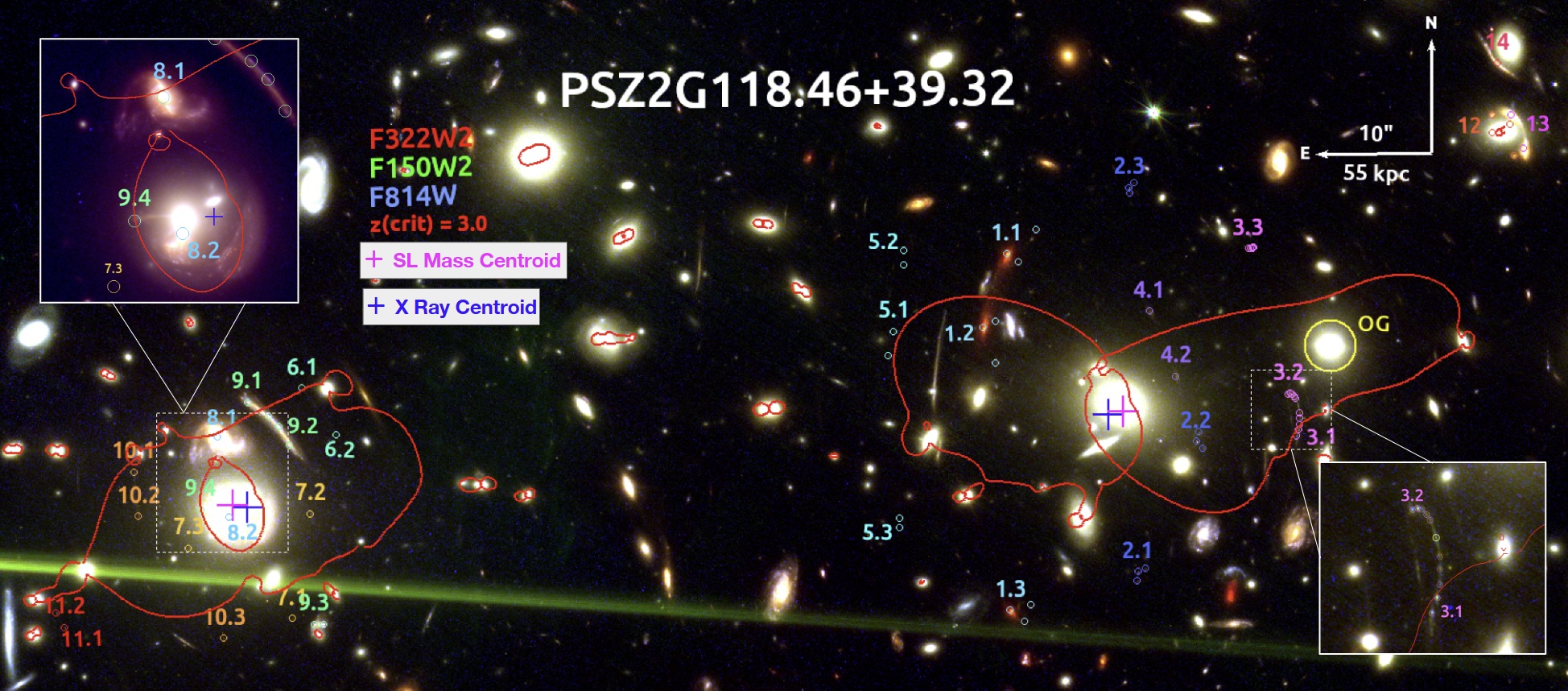}
    \caption{Color composite image of PSZ2\,G118.46+39.32 composed from JWST NIRCam/F322W2 (red), F150W2 (green), and HST ACS/F814W (blue). The lensed galaxies are color coded by system and identified by: System Number.Image Number. The clumps used as constraints in each system are circled in the system color. Circled in yellow in the lower right inset, there is an object that could be a candidate for a highly magnified star, such as those described in \citealt{2024Diego}. Critical curves of the best-fit lens model are plotted in red for sources at z = 3.0. Magenta crosses mark the lensing mass centroids and blue crosses mark the X-ray gas centroids measured from Chandra observation (Obs ID: 11754, PI: Maoz). We located the X-ray centroids with the \textsc{python} package \textsc{pyproffit} \citep{2020Eckert} and the mass centroids with the \textsc{astropy} package \textsc{photoutils}  \citep{larry_bradley_2025_14889440}.}
    \label{fig: lensing model}
\end{figure}

     The model was constructed with 11 multiply imaged systems, decomposed into 30 images with 60 clumps used as strong lensing constraints. The multiple images are labeled in Figure~\ref{fig: lensing model}. Details of the model constraints and best-fit model parameters are given as Data behind the Figure.
     
     Additional strong lensing features appear 40" northwest of the western BCG, labeled as sources 12, 13, and 14 in Figure~\ref{fig: lensing model}. They appear to be due to individual cluster member galaxies. We do not include these in our model because they do not help constrain the overall mass distribution. This could be indicative of additional mass in this region and is worth exploring in future work.

    Spectroscopic redshifts of the lensed galaxies are not available; however, we obtained photometric redshifts for arcs 1.1, 8.1, 9.1/9.2, and 9.3 using the software package \texttt{Prospector} \citep{prospector} with photometry in the available HST and JWST bands. The probability distributions of the estimated photometric redshifts were used to constrain the model redshift priors. The photometric redshift for 1.1 was not used to constrain the model due to large photometric uncertainties.

\section{Results} \label{sec: Results}

    We find that the lens plane is well described by two cluster-scale halos, two free galaxy-scale halos, and 195 fixed galaxy-scale halos. Optimization was performed in the image plane, and the best-fit model has an image plane rms of $0\farcs34$. 
 
    The projected mass density of the cluster is $M(<500\text{kpc})=(4.9 \pm 0.1) \times 10^{14} \Msun$ centered on the west sub-cluster. The masses of each sub-cluster are $M_{\mathrm{West}}(<150\text{kpc}) = (0.82 \pm 0.01)\times 10^{14} \Msun$ and $M_{\mathrm{East}}(<150\text{kpc})=(0.99 \pm 0.01) \times 10^{14} \Msun$, respectively. We find an offset of less than 1.3" between the  X-ray centroids and the lensing mass centroids. This suggests that the cluster is in a pre-merger state, as there is no significant displacement between the X-ray gas and dark matter in clear contrast with major merger events such as the “Bullet Cluster”\citep{Clowe2006}.

    This work represents the first strong lensing model for PSZ2\,G118.46+39.32. Future work may improve the model by obtaining and incorporating spectroscopic redshifts of the lensed galaxies used as constraints. In its current state, the lens model provides a clear result regarding the bimodal distribution of mass within the strong lensing regime of the cluster. 

\begin{acknowledgments}
    This work used observations from JWST-GO-5594 and HST-GO-10491 (DOI:\dataset[10.17909/pwcw-eg72]{https://doi.org/10.17909/pwcw-eg72}). Support for program \#5594 was provided by NASA through a grant from STScI, operated by the AURA under contract NAS 5-03127, and the University of Pittsburgh CRCD, RRID:SCR\_022735, using H2P/MPI cluster (NSF award OAC-2117681). 
\end{acknowledgments}
\onecolumngrid

\twocolumngrid

\software{Astropy \citep{astropy2013},
Photoutils \citep{larry_bradley_2025_14889440},
\texttt{Lenstool} \citep{Jullo2007},
SExtractor \citep{1996A&AS..117..393B}
SAOImageDS9 \citep{Joye2003_ds9},
Prospector \citep{prospector}.}

\bibliographystyle{aasjournal}
\bibliography{references}

\end{document}